\documentclass[12pt]{article}

\usepackage{graphicx}
\usepackage{bm}
\usepackage[affil-it]{authblk}
\usepackage[margin=1in]{geometry}
\usepackage[hidelinks]{hyperref}
\usepackage{siunitx}
\usepackage[T1]{fontenc}
\usepackage{amsmath}
\usepackage{amssymb}
\usepackage[sort&compress]{natbib}
\usepackage{setspace}

\renewcommand{\vec}[1]{\mathbf{#1}}

\date{} 
\title{Subdiffractive confinement of ultrashort mid-IR pulses with photonic funnels}
\author[1]{J. LaMountain\thanks{\texttt{\href{mailto:jacob_lamountain@student.uml.edu}{jacob\_lamountain@student.uml.edu}}}}
\author[2]{A. Raju}
\author[2]{D. Wasserman}
\author[1]{V. A. Podolskiy}

\affil[1]{Department of Physics and Applied Physics, University of Massachusetts Lowell, Lowell, MA 01854, USA}
\affil[2]{The Chandra Family Department of Electrical and Computer Engineering, University of Texas at Austin, Austin, TX 78758, USA}

\begin{document}

\maketitle
\begin{abstract}
	The ability to control the spatial distribution of light, particularly in deep sub-wavelength areas, is important for a range of materials science, microscopy, and communications applications. Separately, materials science and communications rely on the ability to temporally shape the evolution of electromagnetic pulses. In this work we investigate theoretically the propagation of ultrafast pulses inside hyperbolic metamaterials-based photonic funnels, which have been recently used to achieve deep subwavelength (wavelength/30) concentration of monochromatic mid-infrared light. By analyzing the complex spatio-temporal dynamics of the pulse-funnel interaction, we show that  photonic funnels, in general, broaden bandwidth-limited ultrafast Gaussian pulses. We demonstrate that this broadening can be mitigated by pre-chirping the incoming light, realizing simultaneous intensity enhancement and spatio-temporal compression of mid-wave IR light in the all-semiconductor ``designer metal'' funnel platform. Our analysis suggests that, in combination with linear chirp, designer-metal-based photonic funnels can be utilized with \qty{100}{\fs} bandwidth- and diffraction-limited pulses to produce wavelength/30-scale signals of \qty{\sim200}{\fs} duration, with intensity enhancement on the order of 5. Lowering material absorption can further enhance the peak intensity. The results presented can be used to assess the perspectives of ultrafast sub-diffraction light manipulation in other portions of the electromagnetic spectrum by adjusting the (meta)material composition of the funnels.   
\end{abstract}

\newpage
\onehalfspacing
\setcitestyle{super}

 \section{Introduction}
Light emission and absorption occur on the deep subwavelength scale of an individual (quantum) emitter, while light propagation happens on a much larger---wavelength---scale\cite{landau2013electrodynamics}. 
A complete understanding of light's interaction with a localized quantum object requires isolating that object and controlling the incident light, spatially and temporally, to tailor its interaction with the object. Presently, with few exceptions\cite{wagner2014ultrafast,luo2025ultrafast,qin2021ultrafast,novikov2021}, ultra-fast optical pulses are diffraction-limited in space\cite{wang2021,mareev2022}, while sub-diffractive optics\cite{duan2023,lee2021,ma2018} is almost exclusively monochromatic. In this work we analyze theoretically the prospects for realizing ultrafast deep subwavelength optics in the experimentally-viable metamaterial platform.  

The majority of sub-wavelength light manipulation techniques rely on plasmonic or phononic materials that exhibit negative permittivity at the frequency of interest. In multilayered structures, plasmonic modes couple to each other, enabling the unusual hyperbolic response which eliminates the diffraction limit inside the composites\cite{hoffman2007negative,thongrattanasiri2009,hyperlensNarimanov,hyperlensEngheta,liu2007,sun2018}. Conical multilayer-core photonic funnels\cite{Govyadinov2006} based on all-semiconductor ``designer metal'' multilayer hyperbolic metamaterials have been recently shown to confine monochromatic mid-infrared light to deep subwavelength, wavelength/30 areas\cite{evansFunnels,funnels2024}. However, while the spatial control of propagating light in these unique metamaterial structures has been studied, much less effort has been made to understand the temporal evolution of light in these systems. This lack of comprehensive understanding of ultrafast subwavelength light-matter manipulation in composites, including the uncertainty about the relationship between the relative timing of strong confinement and intensity enhancement, motivates the current study. We investigate the evolution of ultrafast pulses within photonic funnels and demonstrate the possibility of achieving, simultaneously, spatial compression of mid-infrared ($\lambda_0\sim\qty{10}{\um}$) pulses to \qty{\sim0.3}{\um}, their temporal modulation at the level of \qty{\sim200}{\fs}, and enhancement of local intensity by a factor on the order of $5$ with realistic materials. Further optimization of performance may be possible. Our results can be applied to spatio-temporal shaping of pulses across the electromagnetic spectrum, from the ultraviolet and visible to the near- and far-infrared domains by using noble-metal\cite{takayama19,chen2024}, transparent conducting oxide-\cite{naik2013}, perovskite-oxide-\cite{bouras2019}, and graphene-\cite{chang2016realization} based hyperbolic metamaterials.

The spatial behavior of light is known to be affected by the diffraction limit: it is impossible to spatially confine propagating waves to the scale that is significantly smaller than the corresponding wavelength. However, when light propagates in uniaxial materials, the effective wavelength of extraordinary (transverse-magnetically-polarized) waves depend on their propagation directions\cite{landau2013electrodynamics}. Mathematically, this wavelength can be related to the wave-vector $\vec{k}$ of the waves propagating in material via $\lambda_{\rm in}=2\pi/|\vec{k}|$. The wavevector, in turn, is connected to the operating frequency $\omega$ and to the dielectric properties of the material (represented by a tensor $\varepsilon$ with (diagonal) components $\varepsilon_{xx}=\varepsilon_{yy}=\varepsilon_\perp,\varepsilon_{zz}\neq\varepsilon_\perp$) by the dispersion relation, 
\begin{equation}
    \frac{k_x^2+k_y^2}{\varepsilon_{zz}}+\frac{k_z^2}{\varepsilon_\perp}=\frac{\omega^2}{c^2},
\end{equation}
with $\omega$ being operating angular frequency and $c$ being the speed of light in vacuum. 

\begin{figure}
    \centering\includegraphics[width=\textwidth]{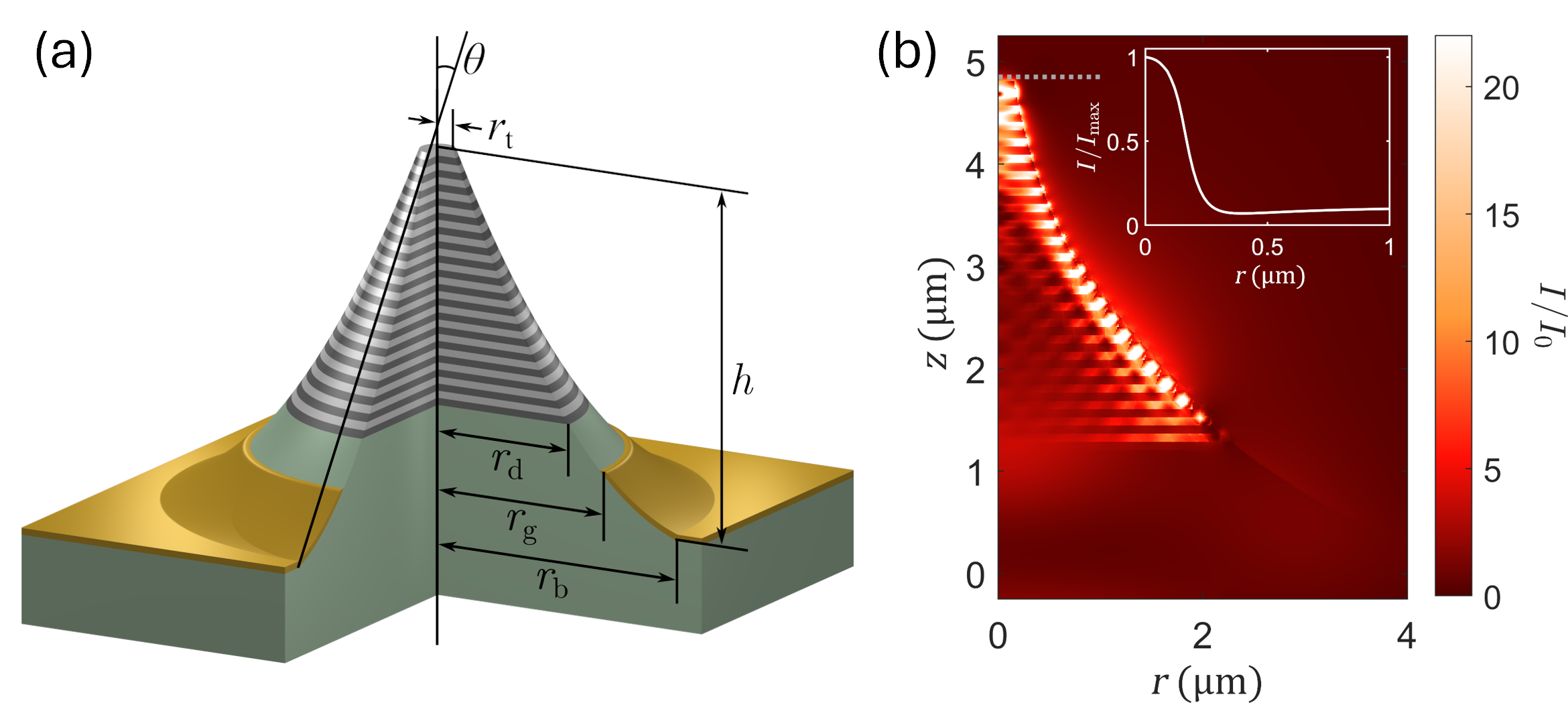}
    \caption{(a) Funnel schematic and (b) Intensity distribution of funnel illuminated from below by monochromatic circularly polarized light; inset shows a normalized trace of intensity across the path indicated by the dotted line, \SI{50}{\nm} above the funnel tip.}
    \label{fig:funnels}
\end{figure}

Materials with such extreme anisotropy that  $\mathrm{Re}(\varepsilon_\perp)\cdot \mathrm{Re}(\varepsilon_{zz})<0$, are known as hyperbolic (meta)materials (HMM) since the wavevectors of the constant-frequency plane waves propagating in such materials lie on the surface of a hyperboloid. Hyperbolic materials support highly oscillatory ($|\vec{k}|\gg \omega/c$) propagating modes. In homogeneous hyperbolic media, these modes propagate preferentially in directions lying on the surface known as a resonance cone\cite{balmain2003,balmain2005}. The angle between this set of preferred propagation directions and the optical axis of the material, known as the critical angle, is given (in the limit of low absorption) by\cite{gan2023}
\begin{equation}
\theta_\mathrm{c}=\arctan\left(\sqrt{-\frac{\mathrm{Re}\left(\varepsilon_\perp\right)}{\mathrm{Re}\left(\varepsilon_{zz}\right)}}\right).
\end{equation}
These highly oscillatory propagating modes form the foundation for light manipulation on a deep subwavelength scale, enabling an increased photonic density of states\cite{narimanovDOS,narimanovTopological,noginov2010controlling,podolskiyEmission,lee2022hyperbolic,takayama19}, anomalous reflection and refraction of light\cite{hoffman2007negative,giles2016,alvarez2022,takayama19}, hyperlenses that use curved hyperbolic media to magnify small objects (or demagnify light for subdiffractive lithography)\cite{hyperlensNarimanov,hyperlensEngheta,liu2007,lee2018,sun2018,takayama19}, hypergratings for coupling and focusing of unstructured and structured light\cite{sreekanth2013,palermo2020hyperbolic,thongrattanasiri2009,li2024nano}, and photonic funnels---conical structures with planar hyperbolic cores that can be used to confine light to, and extract light from, the nanoscale\cite{Govyadinov2006,evansFunnels,funnels2024}. The latter class of structures, illustrated in Fig.~\ref{fig:funnels}, represents the main motivation for this work.

The rest of the manuscript is organized as follows: Section 2 briefly introduces photonic funnels and the designer metal material platform that we use in our analysis; Section 3 describes the formalism we use to relate the frequency and time-domain dynamics; Section 4 presents the main results of this work, the analysis of pulse propagation in the funnels; Section 5 discusses the implications of our analysis for funnels of different shapes and compositions; Section 6 concludes the manuscript. 

\section{Photonic Funnels}

\subsection{Shape and Confinement Scale Metrics}

Photonic funnels are HMM-core conical structures with a base larger than the internal wavelength of light (at the operating wavelength) and a deeply subwavelength tip. A schematic of a photonic funnel is provided in Fig.~\ref{fig:funnels}, which depicts a structure that was previously realized in experiments\cite{funnels2024}. As seen from the figure, the lossless dielectric substrate extends into the core of the funnel, approximately until the cut-off radius of the core, while the gold collar primarily covers the flat part of the sample, preventing the direct transmission of light through the sample, but not covering the sides of the funnel. 

The geometry is parameterized by the funnel height $h$,  base radius $r_\mathrm{b}$, gold collar radius $r_\mathrm{g}$, dielectric radius $r_\mathrm{d}$, and the tip radius $r_\mathrm{t}$. Our previous analysis\cite{funnels2024} demonstrates that the performance of funnels is optimum for conical structures with linear side profile whose angle is close to the critical angle of their HMM core and whose $r_\mathrm{d}$ is close to the cut-off radius for the $\mathrm{EH_{11}}$ mode in the corresponding cylindrical waveguide with homogeneous dielectric core made from the substrate material. However, due to fabrication constraints, the funnels realized in our experiments always have concave sides (seen in Fig.\ref{fig:funnels}). 

Fig.~\ref{fig:funnels}(b) illustrates the typical intensity distribution within the funnel designed to concentrate the incoming diffraction-limited beam. The strong intensity enhancement along the edge of the funnel can be related to the anomalous reflection of light that enters the structure through the base of the funnel, propagates along its optical axis, and is reflected by the core-air interface, which forms an oblique angle with the optical axis of the metamaterial. Note that in properly designed funnels, light is highly concentrated in the tip region of the structure, with the typical confinement scale, $r_\mathrm{eff}$ (defined as a radius where the intensity falls by $1/e^2$), on the order of the tip radius (see Fig.~\ref{fig:funnels}(b) inset).  

Throughout this work we illustrate our results on example of conical funnels; and discuss the implications of these results for the previously-reported experimental realizations of funnels with concave sidewalls \cite{funnels2024} in Section 5. 

\subsection{Materials Response}

One of the most widely used techniques to realize hyperbolic dispersion is to utilize a planar multilayered metamaterial with a bilayer unit cell having one layer with a positive permittivity ($\mathrm{Re}(\varepsilon_\mathrm{d}) > 0$, typically realized with dielectrics) and one with a negative permittivity ($\mathrm{Re}(\varepsilon_\mathrm{m}) < 0$, achievable with either plasmonic or phononic media). When the thicknesses of the two layers are equal and subwavelength ($d_\mathrm{d} = d_\mathrm{m} = d\ll\lambda_0$), as is the case in this work, the effective permittivity of the composite is given by\cite{poddubny2013,takayama19}
\begin{gather}
    \varepsilon_\perp = \frac{1}{2}(\varepsilon_\mathrm{d} + \varepsilon_\mathrm{m})\text{,}\\
    \varepsilon_{zz} = 2\frac{\varepsilon_\mathrm{d}\varepsilon_\mathrm{m}}{\varepsilon_\mathrm{d}+\varepsilon_\mathrm{m}}\text{.}
\end{gather}

In the mid-IR, the required permittivities can be realized with  undoped and highly-doped designer metal semiconductors\cite{Law:12} that provide the benefits of smooth interfaces, excellent thickness control, and monolythic integration that are inherent to molecular beam epitaxy growth. In this work we model the undoped InAlAs and highly doped InGaAs materials realized in previous studies\cite{funnels2024}. We assume that the permittivity of the undoped InAlAs, $\varepsilon_d=10.23$, is independent of operating frequency, $\omega$, while the dielectric response of the highly doped InGaAs layers is adequately  described by  the Drude model,
\begin{equation}
    \varepsilon_\mathrm{m}(\omega) = \varepsilon_\mathrm{b}
    \left(
    1 - \frac{\omega_\mathrm{p}^2}{\omega^2-i\gamma\omega}
    \right)
\end{equation}
with $\varepsilon_\mathrm{b}=12.15$, $\omega_\mathrm{p}=2\pi c/(\qty{12.5}{\um})$, $\gamma=\qty{6.8}{\tera\hertz}$ being the background permittivity, plasma frequency, and scattering rate, respectively. The InP substrate is modeled as having a constant refractive index of $\varepsilon_\mathrm{s}=14.44$. Each HMM layer has a thickness of $d=\qty{80}{\nm}$.

\begin{figure}
\centering\includegraphics[width=\textwidth]{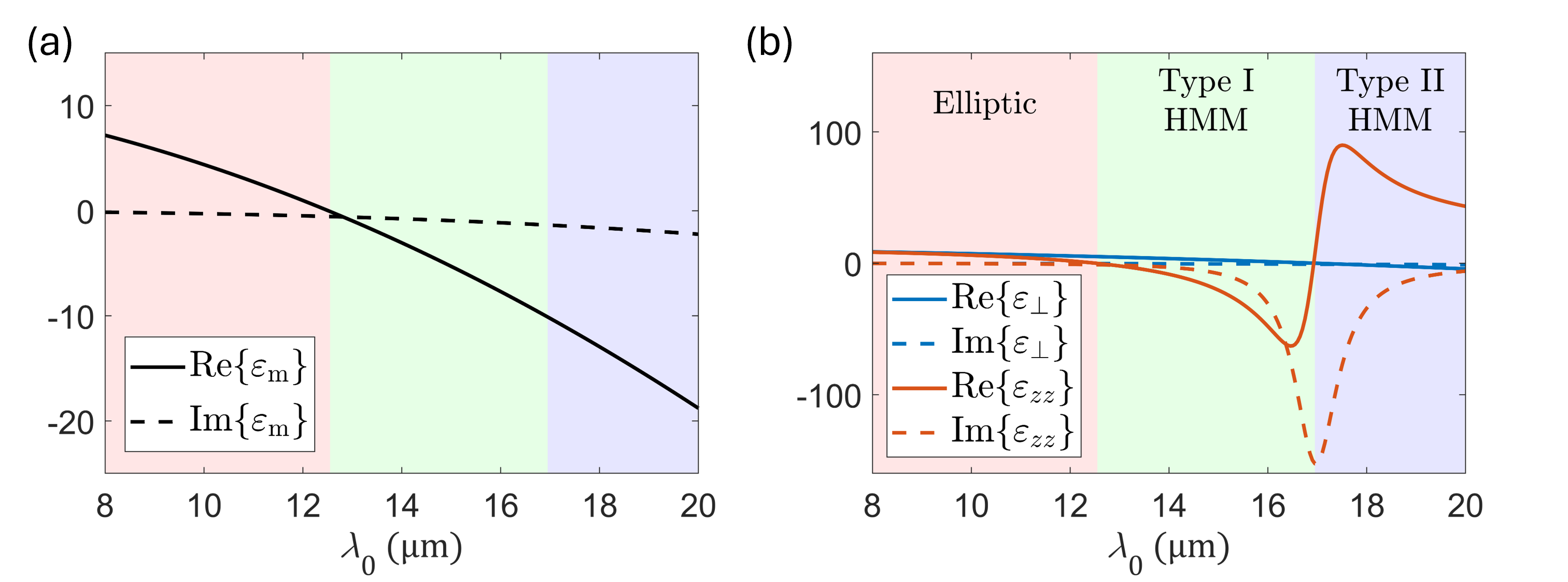}
    \caption{Wavelength dependence of (a) the permittivity of the plasmonic layers and (b) the components of the effective permittivity tensor of the HMM; background color represents elliptic ($\epsilon_\perp,\epsilon_{zz}>0$), type-I hyperbolic ($\epsilon_\perp>0,\epsilon_{zz}<0$) and type-II hyperbolic ($\epsilon_\perp<0,\epsilon_{zz}>0$) frequency ranges, respectively.}
    \label{fig:2}
\end{figure}

The wavelength dependence of the permittivity of the plasmonic layers and of the effective permittivity components of the composite are shown in Fig.~\ref{fig:2}(a) and Fig.~\ref{fig:2}(b) respectively. Of particular note is the type-I hyperbolicity regime, especially the portion of this wavelength range which has relatively low loss, which is optimal for the enhancement and confinement response of monochromatic light within the photonic funnels. For the metamaterials used in this work, this response is centered at \qty{\sim14}{\micro\meter} with a bandwidth of \qty{\sim 3}{\micro\meter}, implying that funnels can support highly confined near fields with durations on the order of \qty{100}{\femto\second}. However, the dispersive nature of photonic funnels, resulting both from the Drude response of the core HMM's plasmonic layers and the finite conical geometry of the funnels\cite{novikov2021,govyadinov2006gain} can lead to significant broadening and re-shaping of the finite-duration pulses. 

\section{Methods}

\subsection{Time and frequency response of chirped Gaussian pulses}
Ultrashort pulses have broad applications in spectroscopy, nano-fabrication, and strong-field physics\cite{pires2015,petersen2018,miyagawa2022}. These pulses are often chirped for reshaping or for the coherent control of energy confinement and dynamic processes\cite{stockman2002coherent,levin2015coherent,weiner2011ultrafast}. A simple model of an ultrashort chirped pulse that is amenable to analytical study is the linearly chirped Gaussian pulse. The time-dependence of the electric field component in such pulses takes the form: 
\begin{equation}\label{eq:chirpt}
    E_\mathrm{c}(t) = E_\mathrm{c} e^{-\left(t/\tau_\mathrm{c}\right)^2} e^{
    i\left(\omega_0t + \frac{1}{2}ut^2\right)}
\end{equation}
where $E_\mathrm{c}$ represents the pulse amplitude, $\tau_\mathrm{c}$ is the pulse duration, $\omega_0$ is the central frequency, and the real-valued parameter $u$ describes the chirp rate (the rate of change of the instantaneous frequency). The full width at half maximum (FWHM) of the intensity distribution in the chirped pulse is given by $T_\mathrm{c}=\tau_c\sqrt{2\ln 2}$\cite{weiner2011ultrafast}. 

It can be shown that the frequency spectrum of the chirped pulse is identical (up to a phase) to that of an unchirped Gaussian pulse of a slightly shorter duration. Specifically, the spectrum of the chirped pulse is given by: 
\begin{equation}\label{eq:chirpw}
    E_\mathrm{c}(\omega)= E_\mathrm{u}\frac{\tau_\mathrm{u} }{\sqrt{2}}\exp\left(-\frac{\tau_\mathrm{u}^2\left(\omega-\omega_0\right)^2}{4}\left(1+i\alpha\right)\right),     
\end{equation}
with $\alpha$ being the real-valued dimensionless chirp parameter quantifying the phase's quadratic frequency dependence that is related to the time-domain chirp rate via
\begin{equation}
    u = \frac{2\alpha}{\tau_\mathrm{u}^2(1+\alpha^2)}\text{,}
\end{equation}
with the subscript $\mathrm{u}$ denoting parameters of the unchirped pulse having the same central frequency and total energy. 

By comparing Eq.(\ref{eq:chirpt}) with the Fourier transform of Eq.(\ref{eq:chirpw}) it is seen that as result of the chirp procedure, the pulse amplitude is reduced by a factor of $(1+\alpha^2)^{-1/4}$ and its duration is increased by a factor of $\sqrt{1+\alpha^2}$ relative to the unchirped source pulse. Note that increasing the magnitude of $\alpha$ does not monotonically increase the magnitude of the chirp rate, which is maximized at $\alpha=\pm1$, but it does increase the magnitude of the group delay dispersion (GDD), which is defined as the second derivative of phase with respect to angular frequency.

Here we parameterize the electric field in these pulses by the dimensionless chirp parameter $\alpha$ and the FWHM of the bandwidth-limited source pulse $T_u$.

\subsection{Simulation of Chirped pulses propagating through photonic funnels}

A finite element method--based model predicting the monochromatic response of  photonic funnels has been developed and described in our previous studies\cite{funnels2024}. The model takes advantage of the cylindrical symmetry of the geometry, and solves for the $(r,z)$-dependence of the electromagnetic fields, assuming harmonic time and angular dependence ($\propto e^{i\left(\omega t-m\phi\right)}$) of all field components, and simulates the field distribution within and around the funnel resulting from  a normally-incident circularly-polarized plane wave. The model takes into account the layered structure of the funnel core, thereby accounting for possible deviations of meta-material response from the predictions of effective medium theory\cite{elser2007}.  

Here, we utilize these monochromatic FEM-based solutions $\vec{E}(\omega,\vec{r})$ to describe the time-dependent evolution of the electromagnetic fields within the funnel via 
\begin{equation}
    \vec{E}(r,z,t) = \int a(\omega) \vec{E}(r,z,\omega)e^{i\omega t}d\omega,
\end{equation}
with $a(\omega)$ being the frequency-dependent complex amplitude that yields the (chirped) Gaussian pulse incident on the funnel. We implement the above integral as a discrete Fourier transform, selecting the spacing of frequencies to appropriately represent the incoming Gaussian pulse and to eliminate artifacts due to the inherently periodic nature of the discrete Fourier transform. 

The performance of photonic funnels is quantified in terms of three figures of merit: the tip intensity enhancement, the confinement radius, and the duration of the signal at the funnel tip. To eliminate the ambiguity related to pulse stretching and amplitude reduction caused by chirping of the incoming pulse, we compare the duration of the pulse at the exit of the funnel to the duration of the bandwidth-limited source of the incoming chirped pulse. Similarly, we define intensity enhancement as the ratio of  the peak intensity \SI{50}{\nano\meter} above the center of the funnel, $I_\mathrm{max}$, to the peak intensity of the source bandwidth-limited pulse above the planar substrate-air interface in the absence of a funnel, $I_0$. 

Notably, both tip intensity and the confinement radius (defined, as above, as the radius at which the intensity  above the funnel tip drops by a factor of $e^2$) are functions of time. Depending on the funnel response and the nature of an incident pulse, the time interval of intensity enhancement may or may not coincide with the interval of subdiffractive light confinement. We will therefore discuss $r_\mathrm{eff}$ as a function of time, or if providing a single value, this will correspond to the confinement radius at the time corresponding to the maximum intensity.

\section{Results}

\begin{figure}
    \centering\includegraphics[width=\textwidth]{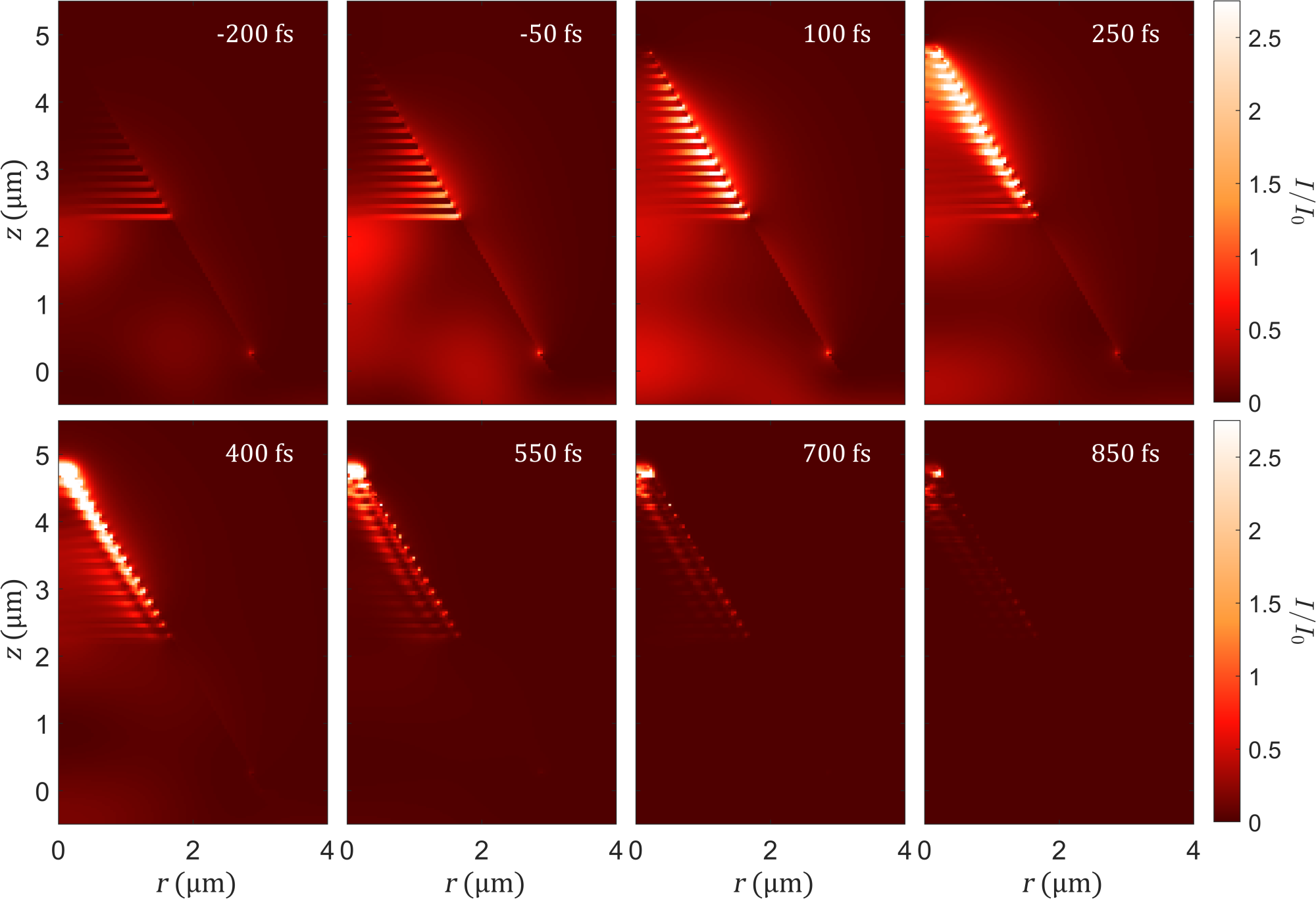}
    \caption{Field intensity as a function of time for the $T_\mathrm{c}=\qty{460}{\fs}$-long pulse with central wavelength $\lambda_0=\qty{14.5}{\um}$ and $\alpha=-4.5$. Anomalous reflection from the oblique HMM-air interface sends highly confined intensity-enhanced critical angle modes towards the funnel tip.}
    \label{fig:snapshots}
\end{figure} 

To gain a comprehensive understanding of the evolution of ultrafast pulses within photonic funnels, we analyze the time-dependent behavior of the electric field within the funnels excited by both unchirped and chirped Gaussian pulses centered at $\lambda_0=\qty{14.5}{\um}$ with pulse duration between \qtylist{50;400}{\fs}. When describing time, we assume that at $t=0$ the maximum of the incident pulse is located \qty{5}{\um} below the funnel. 

The results presented here are for linear funnels with $h=\qty{4.8}{\micro\meter}$, $r_\mathrm{b}=\qty{3}{\micro\meter}$, $r_\mathrm{g}=\qty{2.85}{\micro\meter}$, $r_\mathrm{d}=\qty{1.75}{\um}$ and $r_\mathrm{t}=\qty{150}{\nano\meter}$, with HMM core comprising the above described materials. This configuration yields a funnel angle of \qty{30}{\degree}, corresponding to the HMM's critical angle at \qty{14.2}{\um}, just \qty{100}{\nm} short of the wavelength of maximum tip intensity enhancement. 

The temporal evolution of the intensity within this photonic funnel when excited by a $T_\mathrm{u}=\qty{100}{\fs}$ ($T_\mathrm{c}=\qty{461}{\fs}$)-long pulse with central wavelength $\lambda_0=\qty{14.5}{\um}$ and $\alpha=-4.5$ is illustrated  in Fig.~\ref{fig:snapshots}. The leading edge of the pulse reaches the base of the funnel at around  \qty{-200}{\femto\second}. Between \qty{-50}{\femto\second} and \qty{100}{\femto\second} the intensity along the sidewall of the HMM part of the funnel builds up, reflecting the early onset of the anomalous reflection\cite{giles2016} and the formation of the tightly focused reflected beam. This reflected beam then propagates along the sidewall (that closely matches the direction of the resonance cone within the hyperbolic metamaterial), with intense confined fields forming at the tip by \SI{400}{\femto\second}. As the pulse leaves the funnel, the intensity  throughout the the structure decays over the next couple hundred femtoseconds, with confinement at the tip persisting up until \qty{\sim850}{\fs}. 

\begin{figure}
    \centering\includegraphics[width=\textwidth]{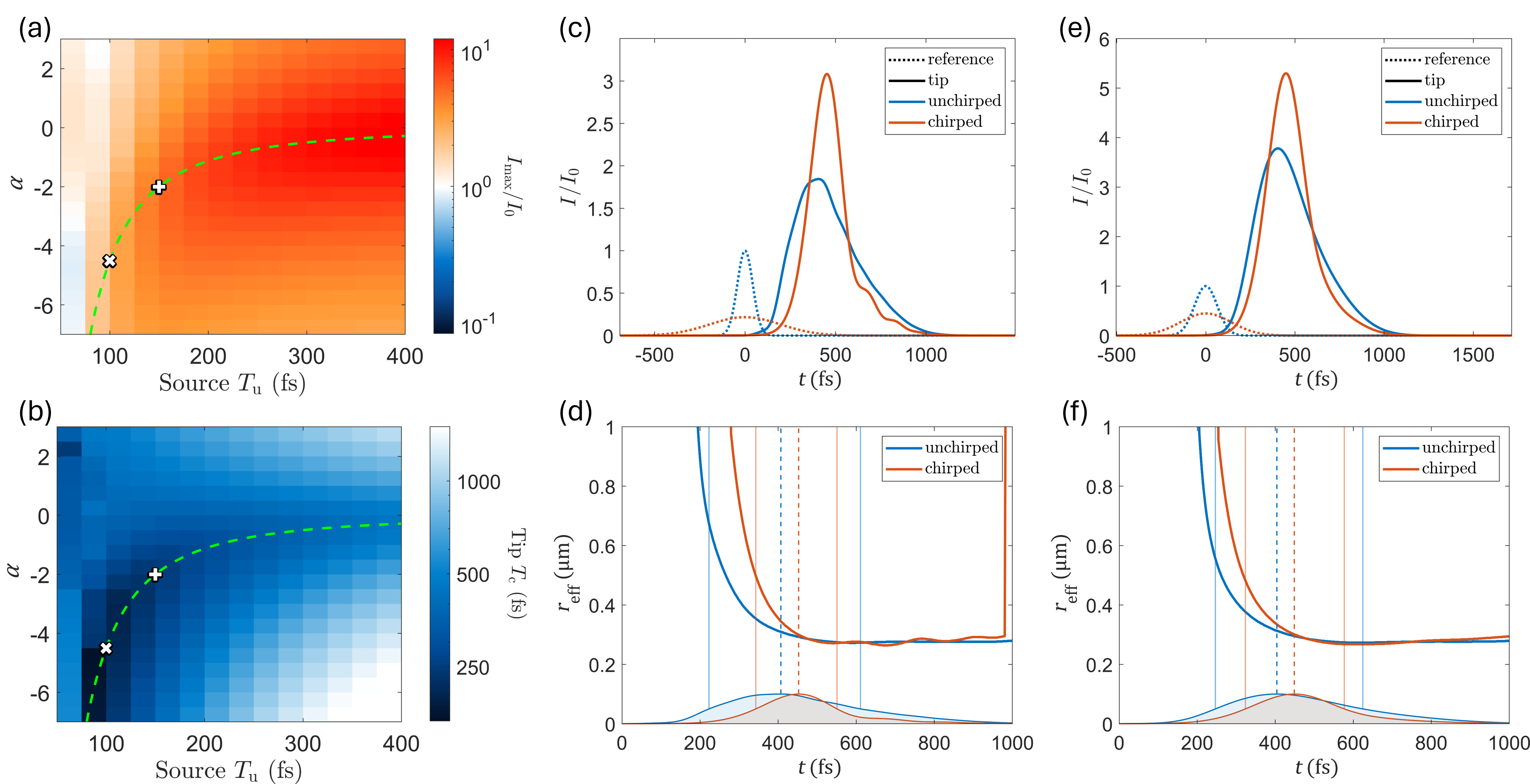}
    \caption{Response of linear-sidewall hybrid funnels to linearly chirped pulses showing (a) intensity enhancement and (b) FWHM at tip as a function of chirp parameter and input duration; green dashed curve sweeps parameters with GDD of \qty{1.6e4}{\fs\squared}; (c,e) intensity above substrate-air interface (dashed curves) and \SI{50}{\nm} above funnel tip (solid curves) and (d,f) confinement radius as functions of time for chirped (orange) and unchirped (blue) source pulses. The results shown in (c,d) and (e,f) are for the pulses marked by $\times$ and $+$ in (a,b), respectively. The vertical lines in (d,f) show times of maximum (dashed) and half-maximum (solid) tip intensity, while the shaded curves show scaled traces of tip intensity.}
    \label{fig:4}
\end{figure}

While the dynamics described above represents the typical behavior of light propagation through the metamaterial-core funnel, the exact time-domain shape of the tightly focused beam strongly depends on the parameters of the funnel as well as on the parameters of the incoming beam. The latter dependence is summarized in Fig.~\ref{fig:4}(a,b), which illustrates the performance of the funnel as functions of the duration and the chirp parameter of the incident pulse. In these simulations we fix the central wavelength of the incident pulse at \qty{14.5}{\um}, maximizing the intensity enhancement in the output pulse in the short-pulse limit. Note that this central wavelength is close to, but greater than, the wavelength of maximum intensity enhancement for monochromatic light (\qty{14.3}{\um}). We attribute this apparent discrepancy to the spectral overlap between the ultrashort pulse and wavelength-dependent intensity response of the funnel (see Ref.\cite{funnels2024} for details). 

As expected, it is seen that funnels excited by bandwidth-limited ($\alpha=0$) Gaussian pulses stretch the incoming beams. For longer pulses ($T_u\gtrsim\qty{300}{\fs}$), whose narrower bandwidths better overlap with the type-I hyperbolic spectral range, responsible for the light-compressing action of the funnel, peak intensities reach the value previously reported for monochromatic analysis. 

Shorter pulses yield smaller intensity enhancement. Ultra-short pulses (with $T_u\lesssim\SI{100}{\fs}$) have spectral bandwidths overlapping the elliptic and type-II hyperbolic regimes and are affected by extreme spatial and spectral reshaping, sometimes accompanied by splitting of the single incident pulse into multiple ``beating'' pulses at the tip. Therefore, the frequency band of the type-I hyperbolic regime imposes a practical limit on the duration of the incident pulse that avoids extreme reshaping. 

Importantly, negative chirping (associated with a positive GDD) is able to compensate for the funnel's broadening, simultaneously shortening the output pulses and enhancing their magnitude (as compared to funnel's response to their unchirped counterparts), demonstrating that the funnel-induced GDD is negative.

For incident source pulses with $\qty{100}{\fs}\lesssim T_\mathrm{u}\lesssim\SI{250}{\fs}$, the minimum tip signal duration occurs for strongly phase modulated pulses with negative chirps ($\alpha<-1$). In this regime the tip field durations are minimized without significant reshaping or beating. Fig.~\ref{fig:4}(c-f) illustrate this behavior in more details by comparing the dynamics of the pre-chirped pulse propagating through the funnel to the dynamics of its bandwidth-limited counterpart. 

For example, when a \qty{100}{\fs} unchirped Gaussian pulse propagates through the funnel, it results in a \qty{390}{\fs} pulse at the tip of the structure, achieving an intensity enhancement of \num{\sim2} (blue lines in Fig.~\ref{fig:4}(c)). The optimal pre-chirping of this pulse, $\alpha=-4.8$, stretches the incoming pulse to \qty{460}{\fs}, while simultaneously decreasing its intensity by a factor of five. However, this pre-chirped pulse yields an intensity enhancement of \num{\sim3} at the funnel tip, while achieving a FWHM duration of \qty{210}{\fs} (orange lines in Fig.~\ref{fig:4}(c)). 

As seen in Fig.~\ref{fig:4}(d), the timing of maximimum tip intensity coincides with significant field confinement in both bandwidth-limited and pre-chirped pulses. Indeed, as the anomalously-reflected beam approaches the tip, the confinement radius quickly drops, reaching $\qty{\sim0.5}{\um}\simeq \lambda_0/30$ at the leading edge of the pulse, and dropping to $\qty{\sim0.3}{\um}\simeq \lambda_0/50$ by the time the maximum intensity is reached, with this confinement continuing for the duration of the pulse. 

Because of their narrower spectrum, longer incident pulses experience larger intensity enhancement, getting somewhat smaller benefits from pre-chirping, with the optimal chirp parameter approaching zero for $T_u\gtrsim 300 fs$ (Fig.~\ref{fig:4}(e)). The temporal dynamics of the confinement radius (Fig.~\ref{fig:4}(f)) remain qualitatively similar to those described above for shorter pulses.

\section{Discussion}

In Section 4 we illustrated time- and space-confinement of light  using, as an example, a  single configuration of the conical funnel and metamaterial core. Here we discuss the implications of our results for other materials and funnel geometries. 

To assess the contributions of material absorption and material dispersion to the dynamics of the highly-confined modes in photonic funnels, the analysis presented above has been repeated for structures whose inelastic loss has been halved as compared with the data presented in Fig.~\ref{fig:2}(a,b). Reducing the inelastic loss simultaneously lowers the absorption and increases the dispersion at the type-I--type-II hyperbolicity transition of the metamaterial. As a result, inelastic loss reduction leads to a narrowing of the low-loss type-I HMM spectral region that underlies the ultrafast dynamics of the highly confined light within the structures. 

\begin{figure}
    \centering\includegraphics[width=\textwidth]{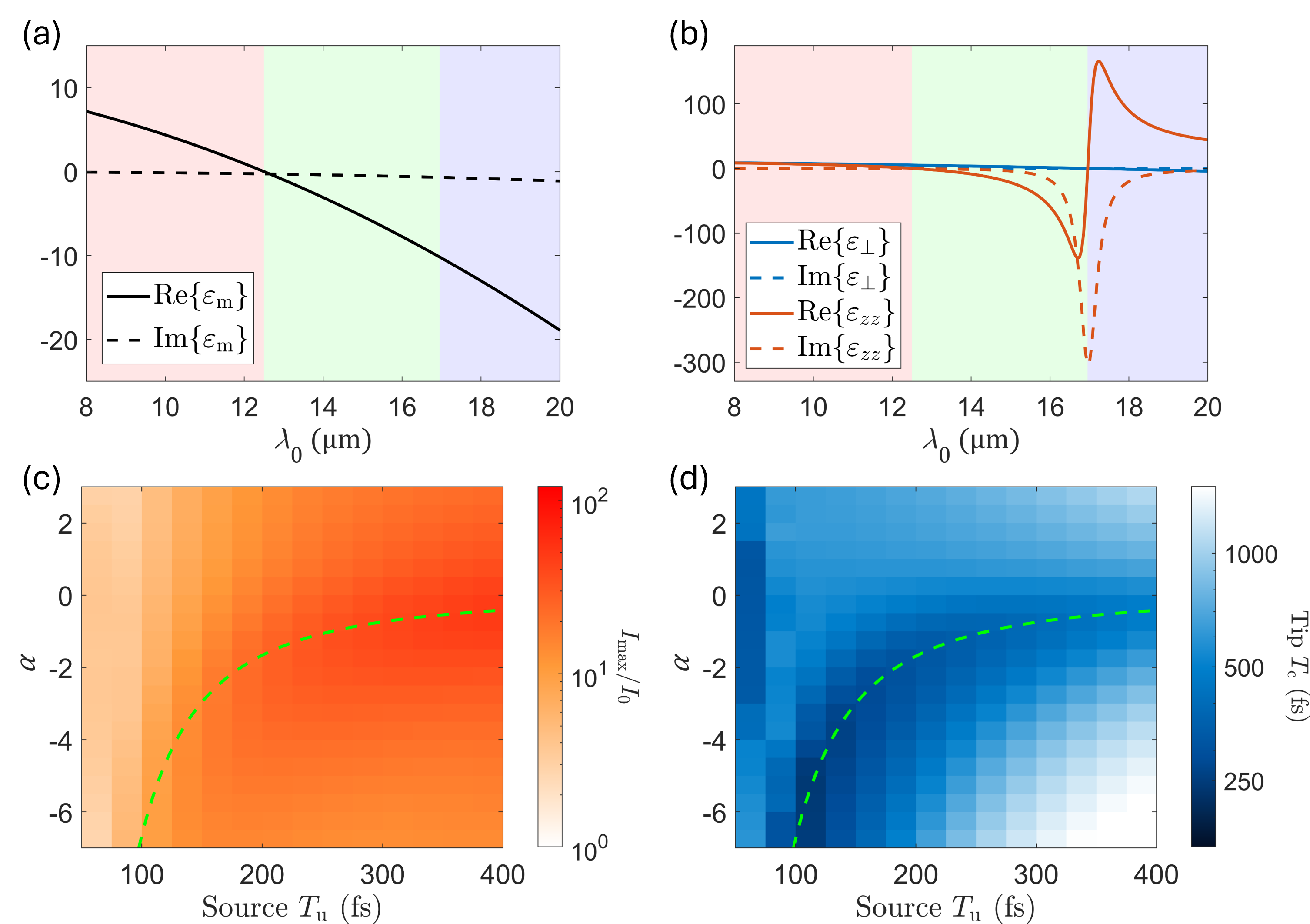}
    \caption{ Materials and funnel response with hypothetical low-loss plasmonic core showing (a) plasmonic layer permittivity; (b) components of the effective permittivity tensor of HMM; (c) intensity enhancement; and (d) FWHM at tip as functions of chirp parameter and source pulse duration. Dashed green lines in (c,d) trace parameters with GDD of  \qty{2.4e4}{\fs\squared}.}
    \label{fig:lowLoss}
\end{figure}

The analysis of temporal reshaping and spatial confinement of light in the low-loss (high dispersion)-based photonic funnels is shown in Fig.~\ref{fig:lowLoss} for pulses with central wavelength of \qty{14.5}{\um}. It is seen that, as a result of loss reduction, the maximum intensity enhancement reaches \num{\sim50} (four times better than what was observed in realistic-loss funnels) while the spectral narrowing of the low-dispersion region increases the duration of the shortest pulse achievable within the funnels to \qty{240}{\fs} (\qty{30}{\fs} longer than the shortest pulse achievable with realistic-loss funnels). Notably, even these short pulses have enhancement factors of about 10, three times those of the highly chirped short pulses in the realistic material. 

Apart from material absorption, the performance of photonic funnels strongly depends on the cross section of the structure. Our previous research has demonstrated that conical funnels with linear cross section whose apex angle is close to the critical angle of the HMM are optimal for achieving spatial confinement and enhancement of the monochromatic light\cite{funnels2024}. At the same time, the wet-etch techniques used to realize photonic funnels in these experiments result in structures that have  concave sidewalls (illustrated in Fig.1). Funnels with concave profiles maintain the same confinement of monochromatic light as their linear-profile counterparts, but with a reduced intensity enhancement.  To analyze the effect of the wall shape on the pulsed performance of the funnels, we repeated the analysis shown in Fig.4 using the parameters of the shape extracted from our previous experimental studies\cite{funnels2024}. Since the shape adjustment increased the wavelength of maximum intensity enhancement, incident pulses with a central wavelength of \SI{14.8}{\micro\meter} were analyzed. The spatial confinement and duration of pulses at the exit of the full-loss concave funnels were comparable to the performance of structures with linear profiles. However, deviation of funnel shape from the optimal linear profile resulted in reduction of the peak intensity of the pulse. The unchirped pulse with \qty{100}{\fs} duration was stretched by the funnel to \qty{\sim400}{\fs}, with intensity enhancement of 0.7. Pre-chirping of this pulse, resulted in the \SI{200}{\femto\second} output with intensity enhancement of \qty{1.2}  with simultaneous deep subwavelength confinement of electromagnetic fields.

The designer metal platform\cite{Law:12,law:14}
enables realizations of the semiconductor-based funnels across the mid-infrared range by properly selecting the doping concentration and funnel geometry for maximal enhancement. As discussed above, doping concentration and metamaterial geometry (concentration of plasmonic layers) should be chosen to minimize material absorption while maximizing the bandwith of low-dispersion type-I hyperbolicity regime. The shape of the funnel should be as close as possible to linear conical structure, with cone angle approaching the critical angle of the metamaterial. 

Given that hyperbolic (meta)materials have been discovered or developed across the electromagnetic spectrum\cite{takayama19,chen2024,naik2013,bouras2019,chang2016realization}, the ultra-fast subwavelength-confined manipulation of intensity should be achievable in hyperbolic-core photonic funnel platforms  from UV to FIR wavelengths.

\section{Conclusions}
We have analyzed the spatial and temporal dynamics of ultrashort pulses propagating through photonic funnels with hyperbolic metamaterial cores.  We have demonstrated that the timing of the highest field enhancement coincides with the timing of sub-wavelength confinement of intensity at the tip of the funnel. We have shown that, in general, the short ($\lesssim 100 fs$) bandwidth-limited pulses are significantly broadened by the funnels due to funnel-induced dispersion. However, this dispersion can be compensated by the negative linear pre-chirping of the incoming pulse. 

The bandwidth of type-I hyperbolicity regime has been identified as a crucial parameter limiting the minimum duration of the sub-diffraction ($\lambda_0/30\ldots \lambda_0/50$)-confined pulse at the tip of the funnel. The intensity of this pulse is limited by material absorption and pulse duration. We have demonstrated that by pre-chirping \SI{100}{\fs} bandwidth- and diffraction-limited pulses, conical funnels with realistic material parameters can be used to produce deeply subwavelength ultrafast \SI{\sim200}{\fs}-signals with intensity enhancement of the order of \numrange{3}{5}, operating at mid-infrared frequency range. Similar duration and confinement (but somewhat smaller enhancement) is expected in the concave funnels realized in recent experiments. 

Time- and space-confinement of optical pulses within the photonic funnels platform can be further optimized across mid-infrared frequency band by adjusting doping and funnel shape within the designer metal platform. Outside the mid-infrared frequencies, the formalism presented in this work can be used to realize ultra-fast and deep subwavelenght confinement with plasmonic\cite{takayama19,chen2024}, transparent conductive oxide\cite{naik2013}, phononic\cite{giles2016}, and other hyperbolic (meta)materials\cite{bouras2019}. 

\section*{Funding}
This research has been supported by the National Science Foundation (awards \#2004298 and \#2004422)
\bibliographystyle{unsrt}
\bibliography{refs}
\end{document}